\documentstyle[preprint,aps,epsfig]{revtex}
\begin{document}
\tightenlines


\title{Collectivity of motion in undercooled liquids and amorphous solids}

\author{H. R. Schober}

\address{Institut f\"ur Festk\"orperforschung, Forschungszentrum J\"ulich,
D--52425 J\"ulich, Germany}

\date{\today}

\maketitle
\begin{abstract} 
The motion of the structure determining components is highly
collective, both in amorphous solids and in undercooled liquids. 
This has been deduced from experimental low temperature data in the tunneling regime
as well as from the vanishing isotope effect in diffusion in 
glasses and undercooled liquids. 
In molecular dynamics simulations of glasses one observes that both 
low frequency resonant vibrations and atomic jumps are centered on 
more than 10 atoms which, in densely packed materials, form chainlike
structures. With increasing temperature the number of atoms jumping
collectively increases. These chains of collectively jumping atoms
are also seen in undercooled liquids. Collectivity only vanishes at higher
temperatures.
This collectivity is 
intimately related to the dynamic heterogeneity which causes 
a non-Gaussianity of the atomic displacements. 
\end{abstract}

\section{Introduction}

Research into the physics of glasses spans a large temperature range,
from the liquid to near $0$~K. There are a number of characteristic
temperatures. The melting temperature,  $T_{\rm m}$, marks the
crossover to the undercooled regime, where the liquid is no longer the 
thermodynamic ground state.
The Kauzmann temperature, $T_{\rm K}$, is the temperature where the entropy 
of the liquid
would fall below the one of the corresponding crystal. It is sometimes
interpreted as an ideal glass transition temperature. 
Before $T_{\rm K}$ is reached upon cooling,
any real system arrests at the glass-transition temperature,
$T_{\rm g}$, into a solid like state, the glass. This temperature $T_{\rm g}$ 
shows 
e.g. as a change in volume expansion or a (smeared) jump in the specific heat.
By definition one has 
in all systems $T_{\rm K}< T_{\rm g} < T_{\rm m}$. The exact value of
the glass-transition temperature depends on cooling rate, pressure  and other
parameters \cite{okamoto:99}.
 
In many materials one observes a sharp upturn of the viscosity
curves at temperatures well above $T_{\rm g}$. This change in the dynamics
of the melt is taken in theories such as the mode coupling theory (MCT) 
\cite{gotze:92} as
the true transition which defines a temperature $T_{\rm c} > T_{\rm g}$.

At the lowest temperatures,
below $\approx 1$~K, the dynamics in glasses is dominated by two-level 
systems \cite{phillips:81} which can be described by the 
tunneling model which was formulated nearly thirty years ago 
\cite{phillips:72,anderson:72}.
In addition to these tunneling states, one observes  
local relaxations  in glasses and at frequencies 
near 1~THz an excess of vibrations compared to the Debye spectrum, given by 
the sound waves. This excess leads to the ``boson peak'' in the 
inelastic scattering intensity. To describe the excess low
energy excitations the tunneling model was extended to the soft potential
model \cite{KKI:83,IKP:87}. From this model ``universal'' 
temperature dependencies
can be derived for temperatures of some 10~K.  Comparing
the model to experiment one finds that the excess low energy excitations
are collective motions of 10 to 100 atoms \cite{BGGS:91,BGGPRS:92}. 
Similar values were found by computer simulations for the low frequency 
resonant modes \cite{LS:91}.

The most popular model to describe the under-cooled liquid is the mode
coupling theory \cite{gotze:92}. In this statistical model a nonlinear
memory kernel leads to a blocking of modes at $T_{\rm c}$. It predicts
scaling relations and dependencies of the form $(T-T_{\rm c})^\gamma$
for quantities such as the diffusion constant. Many aspects of this
theory have been at least semi quantitatively verified by computer
simulations, see e.g. \cite{kob:95a}.   

At present there is no common theory for both the glassy and the super-cooled 
state. The transition from one regime into the other is probed by experiment
and increasingly by computer simulations. The accumulated data should 
help to probe the limits of the existing theories and guide towards a
theory bridging the present gap between glass and liquid theories.

In this contribution we will review computer simulations both in the glass
and in the liquid. The emphasis will be on collective effects observed in 
both.

\section{Simulation models}

Our molecular dynamics simulations are done for systems of 500 to 10000
atoms with periodic boundary conditions. The equations of motion are 
integrated by the velocity Verlet algorithm with a time-step of order fs.
We use cooling rates of $\ge 10^{11}$~K/s and aging times of order ns.
For the model systems of soft spheres and Lennard-Jones atoms these
values correspond to a conversion to Ar.
Zero external pressure was exerted on all but the soft sphere systems
where the volume was kept constant. 

To check the material dependence of the results we
use model systems (soft spheres (SSG) and Lennard-Jones (LJ), both monatomic 
and
binary) and models of the binary
metallic glass CuZr and of Se. Details of the simulations can be
found in the references given further down. 

The soft spheres (SSG) are described by a purely repulsive potential
\begin{equation}
V(R) = \epsilon \left( \sigma / R \right)^n +V_{\rm cut}
\label{eq_SSG}
\end{equation}
where $n=6$ or $n=12$ and $V_{\rm cut}$ is a small correction to give a smooth
cutoff.

As a simple model potential which allows for zero external pressure we
use a Lennard-Jones (LJ) potential
\begin{equation}
V(R) = 4\epsilon \left[ \left( \sigma / R \right)^{12} -  
\left( \sigma / R \right)^{6}  \right] + V_{\rm cut}.
\end{equation}
We study both  monatomic and binary LJ-systems. In the latter 
case we use the parameters of Kob and Anderson \cite{kob:95a} with a different
cutoff.

As a typical example of a binary metallic glass we simulate Cu$_{33}$Zr$_{67}$,
described by a modified embedded atom potential. For 
details see \cite{gaukel:thesis}. The results do not strongly depend on the
choice of interaction. As far as data are available there is good agreement
with the similar NiZr system modeled with a totally different interaction

As an example of a good monatomic glass former we choose selenium. Se
has a coordination number near 2. The amorphous structure consists of
inter-netted rings and chains. We describe it by a potential of the
Stillinger-Weber type. In this potential the covalent bonds are
described by an additional
three body term $V_3\left(|{\bf R}^1 - {\bf R}^2|,|{\bf R}^1 - {\bf R}^3|,
\cos\Theta_{213}\right)$ where $\Theta_{213}$ is the apex angle
\cite{OJRS:96}. The parameters were fitted to both molecular and crystalline
data.

The examples presented in the following will be taken from simulations
of these different materials. We will emphasize the qualitative aspects
which are equal for all these systems.

\section{Vibrations and atomic jumps in the glass}
 
The vibrations in a glass at low temperatures can be most easily studied in
the harmonic approximation. As example we quench samples of up to 5488
atoms 
of a SSG  to $T=0$~K. The
atomic configuration will then correspond to a minimum of the potential 
energy.  We can expand the energy in terms of the displacements from this
minimum energy. The quadratic terms define a dynamical matrix
\begin{equation}
D^{ij}_{\alpha\beta} \ = \ \frac{1}{\sqrt{m_i m_j}}
\frac{\partial^2 \, E_{pot} \, (\{ {\bf R} \} )}{ \partial R_{i,\alpha} \ \partial R_{j,\beta}},
\end{equation}
where $i,j$ denote the atoms and $\alpha ,\beta$ the Cartesian 
coordinates. Diagonalization gives, in harmonic approximation, the frequencies
of the eigenmodes of vibration and their eigenvectors, i.e. their spatial
structure. Fig.~1 shows the vibrational spectrum, $Z(\nu)$, 
together with the corresponding
Debye spectrum calculated from the elastic constants. The area 
between the two curves is due to the excess 
low frequency vibrations, typical for glassy structures. 
In a plot of $Z(\nu )/\nu^2$ one finds a maximum around $\nu = 0.1$,
the boson peak. The eigenmodes at the boson peak frequency have a complicated
structure. At the lowest frequencies they can easily be identified as
resonant (quasi-localized) vibrations. 
They can be decomposed into sound waves and soft local
vibrations \cite{SO:96}. The latter are in full accord with the predictions
of the soft potential model \cite{BGGS:91}. The interaction with the local
vibrations causes an attenuation of the sound waves. 
With increasing frequency  the Ioffe-Regel limit is reached, 
the phonon mean free path drops
to the wavelength. The increasing interaction between the modes leads to
a level repulsion and to $Z(\nu) \propto \nu$. In the $Z(\nu )/\nu^2$
representation, or in the inelastic scattering intensity, this gives a drop
$\propto \nu$ which causes the boson peak maximum.

It is important that the cores of the resonant vibrations are local but
extend over many atoms. Any single atom is stable against isolated
displacements. The atoms in the cores of resonant modes participate also
in high frequency localized vibrations. This is typical for resonant
modes caused by stresses due to configurational ``defects''. They are
not due to some atoms weakly coupled to the rest.

The soft modes 
in glasses originate  from  some atomic configurations where some
direction in the $3N$-dimensional configuration space is locally soft.
This softness should not be confused with the one of the sound-waves which
are soft because all atoms move in phase.
In the extreme case,
the group of atoms forming the center of the soft quasi-local vibration
is  stabilized by the embedding matrix of the rest of the glass in 
a position lying between minima
of the potential energy given by its near neighbors.
To illustrate this, the dashed line in Fig.~2 shows
the average potential energy of the 61 atoms which are most active
in the given mode, ${\langle U(x)\rangle}_{\rm core}$. 
Atoms are considered as active in a given
mode if their amplitude, $\mid {\bf e}^n \mid$,  is at least
30\% of the maximal atomic amplitude in the mode.
The {\it partial} potential energy of these active atoms
is indeed double-well shaped with  minima 
at $x_{\rm m} \approx\pm 1.3$, which 
corresponds to maximal displacements of 
individual atoms by $|{\bf R}^n -{\bf R}_0^n|\approx 0.2-0.3\sigma $ 
from the equilibrium configuration. This maximal atomic displacement is
of the order of the one observed in local low temperature 
relaxations \cite{OS:99}.
   
At finite temperatures one observes these as aperiodic  
transitions from one local configurational minimum
into another. To visualize these
we monitor the total displacement $\Delta R$ from a
starting configuration and define
\begin{equation}
\Delta R(t) = \sqrt{ \sum_n \left({\bf R}^n(t)-{\bf R}^n(0)\right)^2}
\label{delta_r}
\end{equation}
where ${\bf R}^n(t)$ is the position vector of particle $n$ at time $t$ and
${\bf R}^n(0)$ is the one at the starting or reference configuration. 
$\Delta R(t)$ oscillates due to the vibrations and changes
due to relaxations, i.e. due to the transitions from one local energy
minimum to another.
An example is shown in Fig.~3 for the SSG at two temperatures.
At $ T=0.02 T_{\rm g}$ 
the glass clearly jumps between three configurations. Let us denote
the configurations $A,B,C$. The jump sequence is 
$B \to C \to B \to C \to A \to C \to A$.
By quenching to $T=0$~K we find the potential 
energy differences
$\Delta E^B - \Delta E^A = 1.74 \times {10}^{-3}\epsilon$ and
$\Delta E^C - \Delta E^A = 1.81 \times {10}^{-3}\epsilon$. These energy 
differences are of the order of the temperature. The corresponding 
spatial distances between the configurations, Eq.~(\ref{delta_r}),
are $\Delta R^{A-B} = 1.63 \sigma$, $\Delta R^{A-C} = 0.96 \sigma$, and
$\Delta R^{B-C} = 1.0 \sigma$, i.e. they are of the order
of the nearest neighbor distance $R_{NN} = 1.1 \sigma$. The maximal distance
an individual atom travels in these jumps is only $0.3 \sigma$, about
a quarter of  $R_{NN}$. Such relaxations can be observed experimentally
e.g. as telegraph noise in the electric resistivity of point contacts
\cite{kozub:98}.
Increasing the temperature by a factor of four the average displacement
in the initial configuration doubles as expected for a vibrational mean
square displacement. The jumps seen at the lower temperature can no longer
be resolved and new jumps over larger distances are observed.

All observed relaxations are collective jumps localized to 10 or more atoms
forming twisted chain-like structures with some side branching \cite{OS:99}.
An example of such a structure is shown in Fig.~4 \cite{SOL:93}.
The chain structure is a consequence of the dominance of the nearest neighbour 
bond in close packed structures. A low temperature relaxation is only
possible if these bonds are not strongly compressed. 
Using the squared atomic displacement as ``mass'' we find a gyration
radius of about 5 nearest neighbour distances for the chain 
of jumping atoms. The effective dimension
is about 2 as one would expect for a twisted chain with side-branching.
We observe similar
structures in amorphous Se \cite{OS:95,kozub:98} and 
CuZr \cite{gaukel:thesis,OGS:99}. 

In materials with a different inherent structure
these relaxations will be different. In SiO$_2$ we would expect them to
be collective twists of tetrahedra as has been postulated for the soft
vibrations \cite{buchenau:84}. The origin will be, however,
the same, namely
local stresses leading to a softness in one direction of the multidimensional
configuration space. 

Increasing the temperature we observe a marked increase
of the number of atoms participating in a single jump \cite{SGO:97}. 

The local relaxations show two important correlations.
First, in accord with the soft potential model they are 
strongly correlated with the soft vibrations \cite{OS:95,OS:99}. 
Secondly, there is
also a strong but not full correlation between subsequent jumps in the 
same part of the sample.
This latter correlation indicates that one jump can trigger another
one which leads to a correlation in time, i. e. bursts of jumps 
\cite{teichler:01}. It also leads to a
slow increase in time of the number of atoms which have moved 
significantly, the dynamic heterogeneity. We illustrate it in Fig.~5
where we show the mobile atoms at $T\approx 0.15 T_g$ after some ns 
(in Ar units). The effect is again independent of the specific material
studied, whereas the size
of the region 
at a given time and temperature will depend on the particulars of the 
inter-atomic interaction. The atoms which have moved in a given time interval
form a complicated structure. From the time dependence of a suitably
defined correlation function a fractal dimension has been derived for
a binary soft sphere glass \cite{parisi:99}.
The time dependence of the dynamic heterogeneity is reflected also in the
time dependence of the non-Gaussianity discussed further down.

\section{Simulations in the liquid}

Computer simulations in the liquid state, especially with respect to MCT, 
have been discussed in numerous papers and reviews, see e.g. 
\cite{kob:95a,teichler:96,kob:99}. The predictions of MCT are at least
semi-quantitatively reproduced in these calculations. 
We have not done
extensive tests. However, quantities such as the intermediate 
self-scattering function and the self diffusion constant follow the
trends of MCT for both our super-cooled CuZr \cite{GKS:99} and 
Se \cite{CS:00}. Tests of MCT are widely discussed elsewhere in this
conference.

In the liquid state it is no longer straightforward to separate single jumps. 
To
obtain information on the atomic structure of the motion one can study
the difference between structures some ps apart. To remove the effects of
vibrations these configurations have to be averaged over a typical vibrational 
period. Fig.~6 shows an example for under-cooled liquid 
Cu$_{33}$Zr$_{67}$.
There are two striking effects. First, clearly the smaller Cu atoms are
much more mobile. Secondly we observe string like structures similar to
the ones shown for the amorphous materials, Figs.~4 and 5.  
These structures indicate a high degree of collectivity
in the motion also in the under-cooled liquid. They have been
studied extensively for a binary Lennard-Jones system by Donati {\it et al.}
\cite{donati:98}. These authors report a marked increase of collectivity
when the liquid is quenched to $T_{\rm c}$. These mobile strings again
will lead to a dynamic heterogeneity. If one marks the atoms which have moved 
in a given time intervall by further than some cutoff distance, one observes
structures similar to the one shown in Fig.~5 for the glassy state
\cite{muranaka:98}. 
Experimental evidence for
collective particle motion was gained by neutron scattering \cite{russina:00}
its conclusiveness was, however, disputed \cite{schmidt:00}.
Additional evidence is again
the vanishing isotope effect in diffusion \cite{ehmler:98}.

\section{Isotope effect in diffusion}

The isotope effect is the most direct experimental probe of collectivity.
In a monatomic liquid the diffusion constant can 
be written \cite{tyrrell:84}
\begin{equation}
D = D_0 f(T,\rho ) = D^*_0 f(T,\rho ) / \sqrt{m}
\label{eq_Dm}
\end{equation}
where $T$ is the temperature, $\rho$ the atomic density and $m$ is the mass of 
the diffusing particle. 
In the case of different components and different isotopes, considered here,
the situation is more complicated. At low densities 
and high temperatures when diffusion
is dominated  by binary collisions the kinetic approximation should hold
and Eq.~\ref{eq_Dm} should apply approximately for each constituent.
Lowering the temperature or increasing the density effects of collective
motion will gain importance. 

A frequently used measure of this effect is the
isotope effect parameter 
$E$ \cite{schoen:58}
\begin{equation}
E_{\alpha\beta}^\ell = 
\frac{D_\alpha^\ell/D_\beta^\ell -1}{\sqrt{m_\beta^\ell / m_\alpha^\ell} -1}
\label{eq_isotope}
\end{equation}
where the index $\ell$ denotes the different chemical components and 
$\alpha$ and $\beta$ denote different isotopes.
For a tracer atom with a mass of $\overline{m}^\ell + \delta m^\ell$ 
(average mass $\overline{m}^\ell$) the change of the diffusion constant is then
in linear approximation \cite{lantelme:77} 
\begin{equation}
\frac{\Delta D^\ell}{\overline{D^\ell}} 
= - \frac{\delta m^\ell}{2 \overline{m}^\ell} E.
\end{equation} 
An isotope effect parameter of $E_{\alpha\beta}^\ell \approx 1$ indicates
uncorrelated single particle motion whereas collectivity results in 
$E_{\alpha\beta}^\ell \to 0$. 

Formally one can describe the isotope effect by an effective mass
\begin{equation}
(m_\alpha^\ell)_{\rm eff} = m_\alpha^\ell +(N_D^\ell - 1) \overline{m}
\end{equation}
where $N_D$ stands for the effective number
of particles
moving cooperatively. Such a definition can be justified well in the solid
state. There, one has well defined diffusional jumps and $N_D$ is defined
by the multidimensional jump vector. Here we introduce $N_D$ only formally.
The mass dependence of the diffusion constant is then given by
\begin{equation}
D_\alpha^\ell = f^\ell(T,\rho )/ \sqrt{(m_\alpha^\ell)_{\rm eff}} .
\label{eq_D_meff}
\end{equation}
From Eq.~\ref{eq_isotope} one gets then $E_{\alpha\beta}^\ell \approx
1 / N_D^\ell$.

Evidence from earlier measurements of the isotope effect
\cite{tyrrell:84} is conflicting.
Progress was made by simultaneously measuring the diffusion of the
tracer atoms
$^{57}$Co and $^{60}$Co \cite{faupel:90}.
Using this technique for diffusion of Co in
amorphous Co$_{76.7}$Fe$_2$Nb$_{14.3}$B$_7$ a value $E=0.1$ was found
indicating a high degree of collectivity. In contrast for self diffusion 
in crystalline Co one finds $E=0.7$ . There, diffusion is by a
vacancy mechanism which involves essentially single particle jumps with
not too large displacements of the neighbors.
The technique was also applied to a super-cooled melt of
Zr$_{46.7}$Ti$_{8.3}$Cu$_{7.5}$Ni$_{10}$Be$_{27.5}$, and again a very low
isotope effect was observed \cite{ehmler:98}.

We have calculated the isotope effect for a monatomic \cite{KS:00} and 
binary \cite{S:01} Lennard-Jones liquids. Fig.~7 shows the isotope
effect parameters for both components of a binary LJ-liquid, consisting of
80\% large and 20\% small atoms as used in many other simulations, see e.g.
\cite{kob:95a}. 
We find that the values 
are relatively low in the whole temperature 
range investigated, except for the high temperature ones of 
the smaller component. 
The behaviour for the larger atoms closely resembles the
one observed in the mono-atomic LJ-system. These atoms form the backbone of the
structure. The values for the smaller atoms show the same decrease with
temperature but are always clearly higher than those of the 
larger atoms.
This reflects their smaller size. At the highest
temperatures the small atoms seem to move  
through
the matrix of larger atoms by binary collisions only. The observed
high collectivity near $T_c$ is in agreement with the experimental findings of
Ehmler {\it et al.} \cite{ehmler:98} who found $E \approx 0.09$ for
Co diffusion in  super-cooled liquid 
Zr$_{46.7}$Ti$_{8.3}$Cu$_{7.5}$Ni$_{10}$Be$_{27.5}$.
It is similar to the one found in the
monatomic soft sphere glass at low temperatures \cite{OS:99}. The 
Decrease of the isotope effect with lower temperatures  is proportional to
the increase in density. The proportionality factors are, however, different
for the two components. The effect of the density on the diffusion constant
and on $E$ are different, the dynamics clearly depends on density and 
temperature.

The high collectivity both above and
below the glass transition is related to the
dynamic heterogeneity which we expect, therefore, to behave similarly.
An experimentally accessible quantity which probes heterogeneity is 
the non-Gaussianity which we will discuss in the last section.

\section{Dynamic Heterogeneity}

The change of dynamics upon cooling the melt towards the glass transition can
clearly be seen in the self part of the van Hove function  $G^s(r,t)$
which is related to the probability that an atom has moved 
by a distance $r$ during a time $t$:
\begin{equation}
P(r,t) = 4\pi r^2 G^s(r,t) = \left\langle \delta \left( r -
\left| {\bf R}^n(t) - {\bf R}^n(0) \right| \right) \right\rangle.
\label{eq_P}
\end{equation}
At high temperatures $P(r,t)$  is, apart from the geometrical factor $4\pi r^2$ 
nearly perfectly Gaussian and broadens $\propto \sqrt{t}$. 
Upon cooling towards $T_c$,
and beyond, 
a tail to  larger $r$-values grows with time. The self 
correlation function becomes markedly non-Gaussian and  a tail to larger
distances grows with time. Finally approaching $T_c$
additional structure evolves, particularly for the more mobile components,
\cite{roux:89,wahnstrom:91,kob:95a,teichler:97,GKS:99}.  
As example
Fig.~8 shows 
the distribution of the atomic
displacements of Zr$_{67}$Cu$_{33}$ at $T=1000$~K after a
time $t=210$~ps, i.e. during the early time of the so called 
$\alpha$-relaxation. The second peak grows at a fixed, time independent,
position roughly equal to the mean nearest neighbor distance.
From this structure of $P(r,t)$ one can conclude that there are 
preferred positions on the relevant time scales. This does, however, not
necessarily mean that these positions are reached in a single jump,
as frequently assumed.    
From an analysis of the evolution with time it has been concluded that
the evolution of this structure in $P(r,t)$ is accompanied with a strong
increase of back correlation \cite{GS:98}.

This strong deviation of $P(r,t)$ indicates that some atoms have been much
more mobile than the average. This effect is called dynamic heterogeneity.  
To quantify it the
non-Gaussianity parameter \cite{rahman:64} is often used
\begin{equation}
\label{eq_ngp}
\alpha_2(t)=\frac{3 <\Delta r^4(t)>}{5 <\Delta r^2(t)>^2}-1,
\end{equation}
where $<...>$ denotes time averaging, $\Delta r^2(t)$ is the mean
square
displacement and $\Delta r^4(t)$ is the mean quartic displacement.
This parameter is defined  so that it is equal to zero when the atomic
motion is homogeneous. Experimentally it can be obtained from the
$q$-dependence of the Debye-Waller factor \cite{zorn:97}. It has also
been calculated in numerous molecular dynamics simulations
of liquids, e.g. by Kob {\it et al.}
for the binary Lennard-Jones system \cite{kob:95a,kob:97}. The limiting
values for both times $t=0$ and $t=\infty$ is $\alpha_2 = 0$. 
The latter limit reflects the ergodicity of the system for long times.
Starting 
from $t=0$, $\alpha_2(t)$ rises in general monotonically to a maximum from 
where it drops again monotonically. The maximum value is around 0.2
in the hot liquid and rises strongly in the under-cooled liquid, where
a maximum value of 3 has been reported \cite{kob:95a}. The position of
the maximum in time is at high temperatures in the ps range and correlates
in the super-cooled liquid with the onset of the $\alpha$-relaxation
which is attributed to long range motion. This general behavior is
observed in all systems we have studied.

In the previous sections we have seen that on an atomic level the
dynamics in the glass and the under-cooled liquids are very similar. To
see this quantitatively for the heterogeneity we have calculated 
$\alpha_2(t)$ for different temperatures above and below the glass 
transition. In Fig.~8 we show this for both components of the binary
LJ-system in a log-log
representation of $t \cdot \alpha_2(t)$ versus time. First we see the
general trends discussed above. These hold not only in the liquid
but also in the glass. Secondly we observe an asymptotic linear 
increase of $\alpha_2(t) \propto \sqrt{t}$. For the lower temperatures
this behaviour stretches over several ns
(when one converts the LJ-units to ones appropriate for Ar). 
This dependence can be
explained by the above mentioned strong correlation between successive
collective jumps \cite{CMS:00}. One can envisage a chain of atoms jumping. This 
jump can either be reversed in a successive jump or might trigger
a jump of another chain which will involve many but not all atoms
which have jumped previously.

\section{Conclusion}
We have shown by computer simulations of different materials that motion
both in the glass and in the under-cooled liquid is highly collective. The
dynamics in the glass at low temperatures can be described by the soft
potential model which postulates similar structures for tunneling centers,
quasi localized vibrations and local relaxations. These local motions
involve groups of ten and more atoms forming in dense packed systems
predominantly chain structures. These structures are closely related
to resonant vibrations which cause the boson peak in the inelastic
scattering intensity. Successive jumps in the same part of the sample are
correlated. 
With increasing temperature the
number of atoms jumping collectively increases in the amorphous state.
Similar collectivity is also
observed in the undercooled liquid where it increases upon quenching.
The decrease of the isotope effect of the
diffusion upon quenching to the glass transition indicates again an
increase in collectivity, in agreement with experiment and other simulations.
The collectivity of motion is
related to the dynamic heterogeneity.
The non-Gaussianity parameter, a measure of the dynamic heterogeneity,
increases rapidly in the undercooled liquid and varies smoothly through 
the glass transition.

\section{Acknowledgment}
The author acknowledges the essential contributions by 
his coworkers B. B. Laird, C. Oligschleger,
C. Gaukel, M. Kluge, D. Caprion and V. L. Luchnikov.
This work was partially supported by the Deutsche Forschungsgemeinschaft in the
Schwerpunkt ``Unterk\"uhlte Metallschmelzen: Phasenselektion und
Glasbildung''.
We also acknowledge the financial support by the 
A. von Humboldt foundation.


\begin{thebibliography}{10}

\bibitem{okamoto:99}
P.~R. Okamoto, N.~Q. Lam, and L.~E. Rehn, {\em Solid State Physics}, edited by
  H. Ehrenreich and F. Saepen (Academic Press, San Diego, 1999), Vol.~52, p.\
  2.

\bibitem{gotze:92}
W. G\"otze and A. Sj\"olander, Rep. Prog. Phys {\bf 55}, (1992) 241.

\bibitem{phillips:81}
{\em Amorphous Solids: Low Temperature Properties}, edited by W.~A. Phillips
  (Springer-Verlag, Berlin, 1981).

\bibitem{phillips:72}
W. Phillips, J. Low Temp.\ Phys. {\bf 7},  (1972) 351.

\bibitem{anderson:72}
P.~W. Anderson, B.~I. Halperin, and C.~M. Varma, Philos. Mag. {\bf 25}, 
  (1972) 1.

\bibitem{KKI:83}
V.~G. Karpov, M.~I. Klinger, and F.~N. Ignatiev, Sov. Phys. JETP {\bf 57}, 
   (1983) 439.

\bibitem{IKP:87}
M.~A. Il'in, V.~G. Karpov, and D.~A. Parshin, Sov. Phys. JETP {\bf 65},
  (1983) 165.

\bibitem{BGGS:91}
U. Buchenau, Y.~M. Galperin, V.~L. Gurevich, and H.~R. Schober, Phys. Rev. B
  {\bf 44}, (1991) 5093.

\bibitem{BGGPRS:92}
U. Buchenau {\it et~al.}, Phys. Rev. B {\bf 46}, (1992) 2798.

\bibitem{LS:91}
B. B. Laird and H. R. Schober, Phys. Rev. Lett {\bf 66}, (1991) 636;
H.R. Schober and B. B. Laird, Phys. Rev. B {\bf 44} (1991) 6746. 

\bibitem{kob:95a}
W. Kob and H. Andersen, Phys. Rev. E {\bf 51},   (1995) 4626; {\it ibid.}
{\bf 52}, (1995) 4134.

\bibitem{gaukel:thesis}
C. Gaukel, Berichte des Forschungszentrums J\"ulich {\bf 3556},    (1998).

\bibitem{OJRS:96}
C. Oligschleger, R.~O. Jones, S.~M. Reimann, and H.~R. Schober, Phys. Rev. B
  {\bf 53},  6165  (1996).

\bibitem{SO:96}
H.~R. Schober and C. Oligschleger, Phys. Rev. B {\bf 53},  11469  (1996).

\bibitem{LMNS:00}
V. A. Luchnikov, N. N. Medvedev, Yu. I. Naberukhin and H. R. Schober,
Phys. Rev. B {\bf 62} (2000) 3181.

\bibitem{OS:99}
C. Oligschleger and H.~R. Schober, Phys. Rev. B {\bf 59},  811  (1999).

\bibitem{kozub:98}
V.~I. Kozub and C. Oligschleger, J. Phys.: Condens. Matter {\bf 10},  8033
  (1998).

\bibitem{SOL:93}
H.~R. Schober, C. Oligschleger, and B.~B. Laird, J. Non-Cryst. Sol. {\bf 156},
  965  (1993).

\bibitem{OS:95}
C. Oligschleger and H.~R. Schober, Solid.\ State Commun. {\bf 93},  1031
  (1995).

\bibitem{OGS:99}
C. Oligschleger, C. Gaukel, and H.~R. Schober, J. Non-Cryst. Sol. {\bf
  250-252},  660  (1999).

\bibitem{buchenau:84}
U. Buchenau, N. N\"ucker, and A.~J. Dianoux, Phys.\ Rev.\ Lett. {\bf 53},  2316
   (1984).

\bibitem{SGO:97}
H.~R. Schober, C. Gaukel, and C. Oligschleger, Defect and Diffusion Forum {\bf
  143-147},  723  (1997).

\bibitem{teichler:01} 
H. Teichler, J. Non-Cryst. Solids in print.

\bibitem{parisi:99}
G. Parisi, J. Phys. Chem. B {\bf 103}, 4128 (1999).

\bibitem{teichler:96}
H. Teichler, Phys.\ Rev.\ Lett. {\bf 76},  62  (1996).

\bibitem{kob:99}
W. Kob, J. Phys.: Condens Matter {\bf 11},  R85  (1999).

\bibitem{GKS:99}
C. Gaukel, M. Kluge, and H.~R. Schober, J. Non-Cryst. Sol. {\bf 250-252},  664
  (1999).

\bibitem{CS:00}
D. Caprion and H.~R. Schober, Phys. Rev. B  {\bf 62} (2000) 3709.

\bibitem{donati:98}
C. Donati {\it et~al.}, Phys. Rev. Lett. {\bf 80},  2338  (1998).

\bibitem{muranaka:98}
T. Muranaka and Y. Hiwatari, J. Phys. Soc. Japan {\bf 67}, 1982 (1998).

\bibitem{russina:00}
M. Russina, F. Mezei, R. Lechner, S. Longeville, and B. Urban, Phys. Rev.
Lett {\bf 84}, 3630 (2000).

\bibitem{schmidt:00} 
W. Schmidt, M. Ohl, and U. Buchenau, Phys. Rev. Lett. {\bf 85}, 5669 (2000);
M. Russina {\it et al.} {\it ibid} p. 5670.
 
\bibitem{ehmler:98}
H. Ehmler {\it et~al.}, Phys.\ Rev.\ Lett. {\bf 80},  4919  (1998).

\bibitem{tyrrell:84}
H.~J.~V. Tyrrell and K.~R. Harris, {\em Diffusion in Liquids} (Butterworth,
  London, 1984).

\bibitem{schoen:58}
A.~H. Schoen, Phys.\ Rev.\ Lett. {\bf 1},  138  (1958).

\bibitem{lantelme:77}
F. Lantelme, P. Turq, and P. Schofield, J. Chem. Phys {\bf 67} (1977) 3869.

\bibitem{faupel:90}
F. Faupel, P.~W. H\"uppe, and K. R\"atzke, Phys.\ Rev.\ Lett. {\bf 65},  1219
  (1990).

\bibitem{KS:00}
M. Kluge and H.~R. Schober, Phys. Rev. E  in print  (2000).

\bibitem{S:01}
H. R. Schober, Solid State Commun. {\bf 119} (2001) 73.

\bibitem{roux:89}
J. N. Roux, J. L. Barrat, and J. P. Hansen, Phys. Rev. A {\bf 38} (1989) 454.

\bibitem{wahnstrom:91}
G. Wahnstr{\o}m, Phys. Rev. A {\bf 44} (1991) 3752.

\bibitem{teichler:97}
H. Teichler, Defect and Diffusion Forum {\bf 143-147} (1997) 717.

\bibitem{GS:98}
C. Gaukel and H. R. Schober, Solid State Commun. {\bf 107} (1998) 1. 

\bibitem{rahman:64}
A. Rahman, Phys. Rev. {\bf 136},  A405  (1964).

\bibitem{zorn:97}
R. Zorn, Phys.\ Rev.\ B {\bf 55},  6249  (1997).

\bibitem{kob:97}
W. Kob {\it et~al.}, Phys. Rev. Lett. {\bf 79},  2827  (1997).

\bibitem{CMS:00}
D. Caprion, J. Matsui, and H. R. Schober, Phys. Rev. Lett {\bf 85} (2000) 4293. 

\end{thebibliography}

\eject
\begin{figure}
  \epsfig{figure=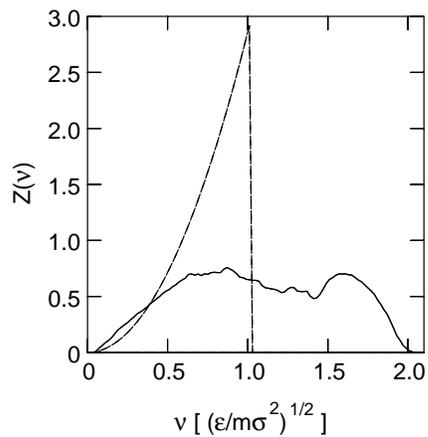,width=5.8cm}
\label{fig_spectrum}
\caption{Configurationally averaged vibrational density of states of the
soft sphere glass (solid line) with $n=6$ and Debye spectrum (dashed line)
\protect\cite{SO:96}}.
\end{figure}
\begin{figure}
\epsfig{figure=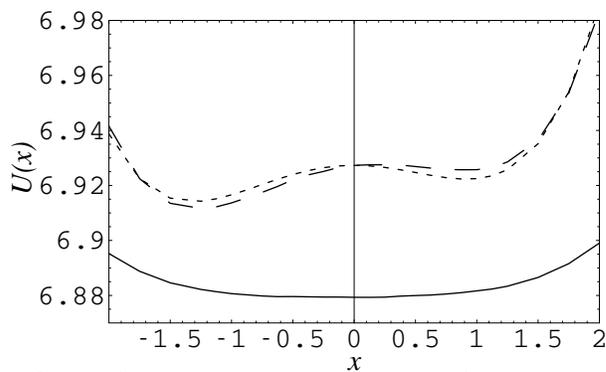,width=8cm,angle=0}
\caption{Average potential energy of atoms in a single soft vibrational 
mode with
frequency $\nu=0.0985$, participation ratio $p=0.23$. 
Solid line: potential energy
averaged over all atoms in the system. Dashed line:
partial 
potential energy, averaged over 61 atoms of the core of the mode.
Dotted line: least squares fit of the partial potential energy
by a soft potential polynomial
\protect\cite{LMNS:00}. The average nearest neighbour distance is $1.1 \sigma$.}
\end{figure}
\eject
\begin{figure}
\epsfig{figure=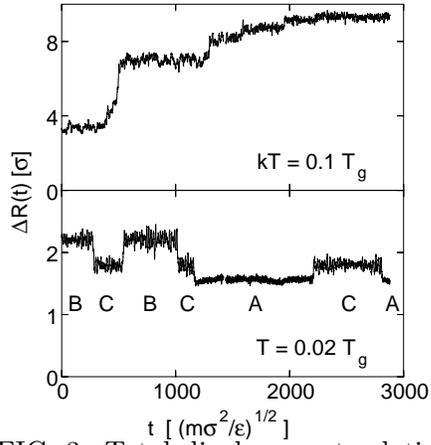,width=5.8cm}
\label{fig_pir}
\caption{Total displacement relative to a local minimum configuration as
function of time for one sample of the glass of Fig.~1 
with $N=5488$ at two temperatures. Please note the different scales for 
\protect$\Delta R$ \protect\cite{OS:99}.}
\end{figure}

\begin{figure}
  \epsfig{figure=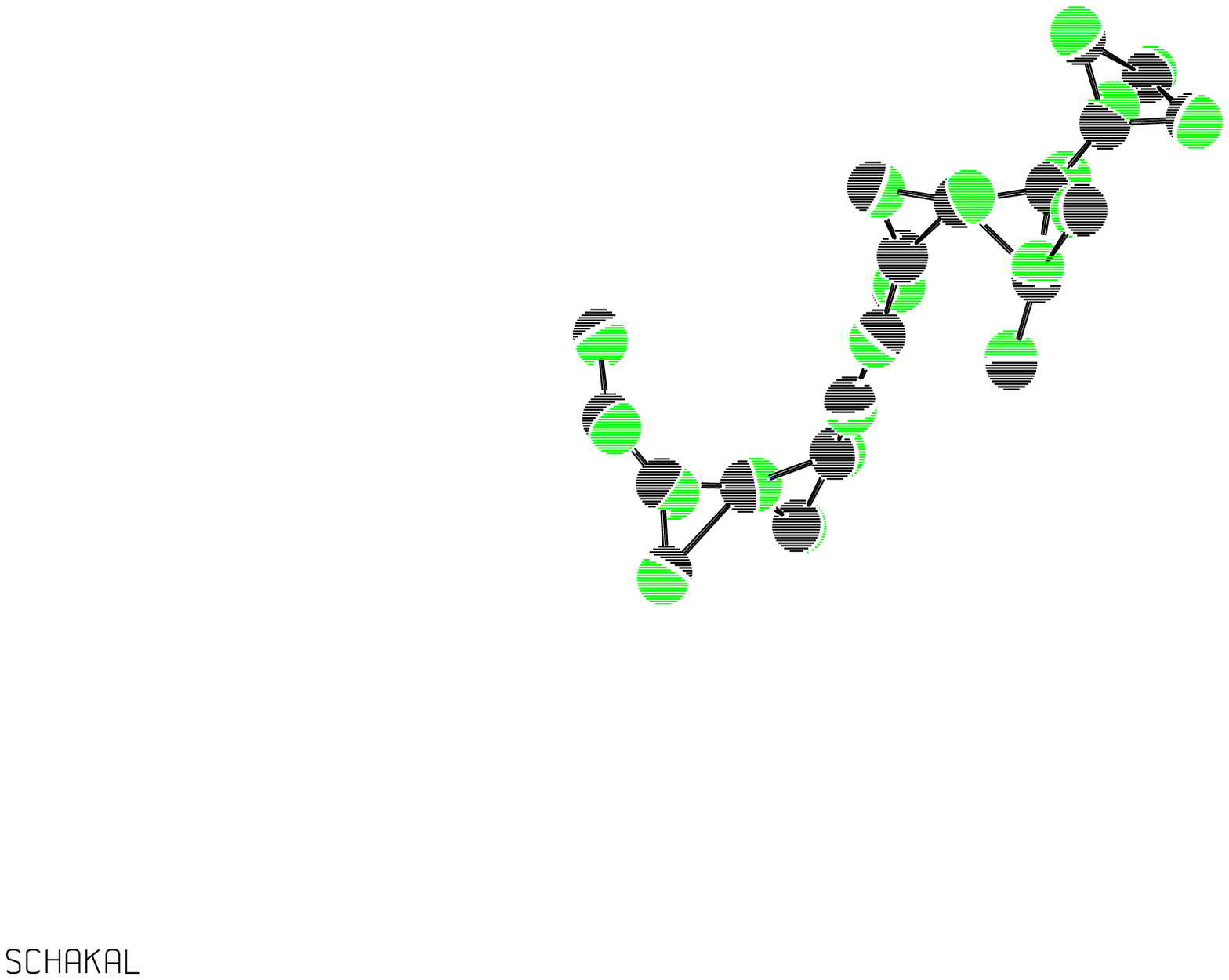,width=5.8cm}
\label{fig_jump}
\caption{Collective jump in the glass of Fig.~1
at $T=0.15T_{\rm g}$.
The initial positions of the atoms are shown by the dark spheres and the
final ones by the shaded spheres. The bonds connect nearest neighbours.
Shown are all atoms with more than 40\% of the maximal atomic displacement
\protect\cite{SOL:93}.}
\end{figure}
\eject
\begin{figure}
\epsfig{figure=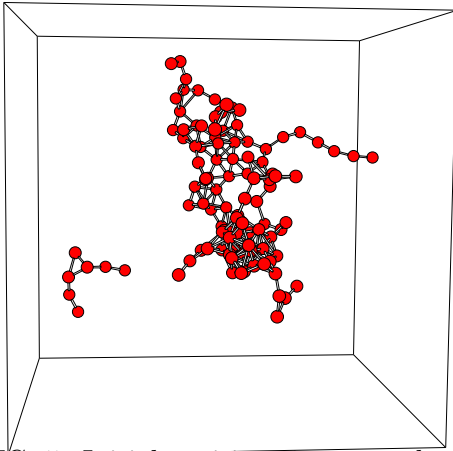,width=5.8cm}
\label{fig_hetero}
\caption{Initial positions in a sample of the glasses of atoms
displaced by more than 0.3 of the average
nearest neighbour distance during  a time interval $\Delta t =
3600(m\sigma^2 /\epsilon )^{1/2}$ at $T \approx 0.15 T_{\rm g}$. 
The total displacement is approximately 8 nearest neighbour distances
\protect\cite{OS:99}.}
\end{figure}

\begin{figure}
\epsfig{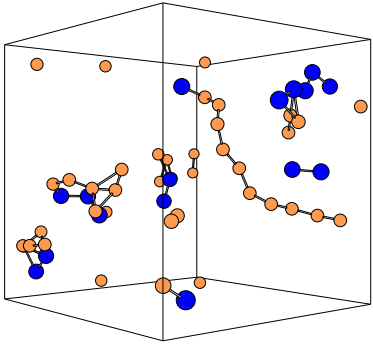}
\label{fig_CuZr}
\caption{Atoms in a melt of Cu$_{33}$Zr$_{67}$ at $T=1200$~K
with the largest displacements between time averaged  
configurations separated by 6.5~ps. 
Shown are Cu (light spheres)
and Zr (dark spheres) atoms displaced by more than 1.6\AA\ and
1.45\AA , respectively \protect\cite{SGO:97}.}
\end{figure}
\eject

\begin{figure}
\epsfig{figure=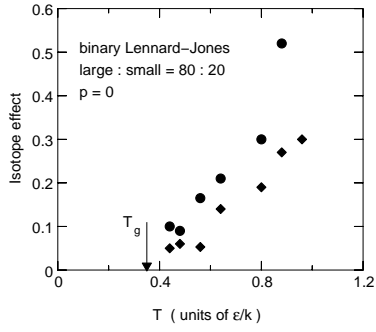,width=5.8cm}
\label{fig_isotope}
\caption{Isotope effect in a binary LS-liquid as function of temperature at 
equilibrium density.
The diamonds and circles refer to the larger and smaller atoms, respectively. 
The arrow indicates
the glass transition temperature \protect\cite{S:01}.}
\end{figure}

\begin{figure}
\epsfig{figure=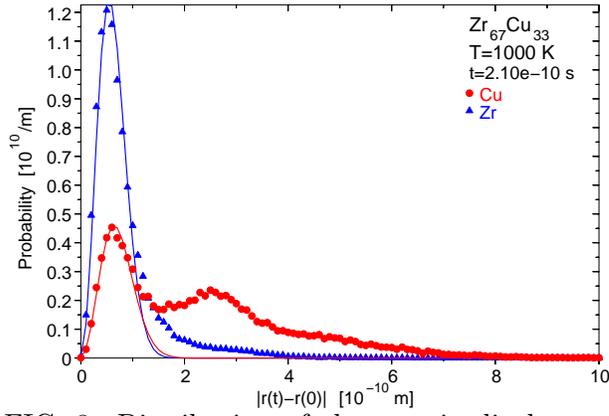,width=8cm,angle=0}
\caption{Distribution of the atomic displacements $P(r,t)$ in Cu$_{33}$Zr$_{67}$ near $T_c$ after 210~ps, 
calculated in  MD (symbols). The line shows a Gauss-fit to the small distance
part \protect\cite{gaukel:thesis}.}
\end{figure}
\eject
\begin{figure}
\epsfig{figure=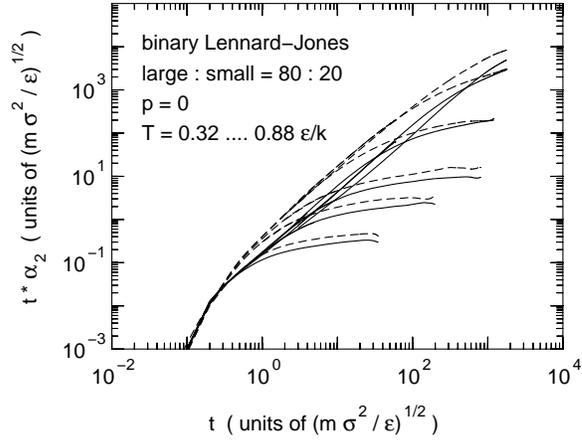,width=8cm,angle=0}
\label{fig_nongauss}
\caption{Non-Gaussianity parameter multiplied by time  
for the binary LJ-system of Fig.~7 above and below the glass transition 
temperature. The solid and dashed lines refer to the larger majority 
and smaller minority components, respectively.
The curves refer to the temperatures  
(from
bottom to top): 0.88, 0.56, 0.48, 0.40, 0.36 and 0.32 in units of 
$\epsilon / k$.}
\end{figure}

\end{document}